\documentclass[12pt]{article}
\usepackage[margin=1in]{geometry}
\usepackage{amsmath}
\usepackage{times}
\usepackage{graphicx}
\usepackage{color}
\usepackage{rotating}
\usepackage{bbm}
\usepackage{latexsym}

\usepackage{epsfig}
\usepackage{amssymb}
\usepackage{algorithm}
\usepackage{algorithmic}
\usepackage{slashbox}

\begin{document}
\begin{center} \textbf{ \Large Estimating an Inverse Gamma distribution} \end{center} 
\begin{center} A. Llera, C. F. Beckmann.\end{center} 
\begin{center} {\footnotesize Technical report \\ Radboud University Nijmegen \\ Donders Institute for Brain Cognition and Behaviour.} \end{center} 
\vspace{0.5 cm}
\begin{center} {\bf Abstract} \end{center}
In this paper we introduce five different algorithms based on method of moments, maximum likelihood and full Bayesian estimation for learning the parameters of the Inverse Gamma distribution. We also provide an expression for the KL divergence for Inverse Gamma distributions which allows us to quantify the estimation accuracy of each of the algorithms. All the presented algorithms are novel. The most relevant novelties include the first conjugate prior for the Inverse Gamma shape parameter which allows analytical Bayesian inference, and two very fast algorithms, a maximum likelihood and a Bayesian one, both based on likelihood approximation. In order to compute expectations under the proposed distributions we use the Laplace approximation. The introduction of these novel Bayesian estimators  opens the possibility of including Inverse Gamma distributions into more complex Bayesian structures, e.g. variational Bayesian mixture models. The algorithms introduced in this paper are computationally compared using synthetic data and interesting relationships between the maximum likelihood and the Bayesian approaches are derived.

\vspace{0.5 cm}
\section{Introduction}

The Inverse Gamma distribution belongs to the exponential family and has positive support. In most cases, the Gamma distribution is the one considered for modeling positive data \cite{Lit:Gm_2,Lit:Gm_1,minka2002estimating,Lit:Gm-N-Gm_1}, and the Inverse Gamma remains marginally studied and used in practice. An important structural difference between these two distributions is that while the Inverse Gamma mode is always positive the Gamma one can be zero; this fact makes the Inverse Gamma very attractive to for example distinguish some kind of positive ’activation’ from stochastic noise usually modeled using a Gaussian distribution \cite{Lit:invg1,Lit:invg2}. In such scenario, the advantage with respect to the Gamma distribution can be specially important for cases with low signal to noise ratio since modeling the activation as Gamma distributed can provide a exponential decay fit to the activation \cite{Llera_ohbm_2015}.  
This slight but important difference from the Gamma distribution as well as the absence of analytical expressions for the estimation of all its parameters motivates our search for efficient algorithms to estimate inverse gamma distributions. 
The most traditional approach considered to estimate the parameters of a distribution is the Method of Moments (MM). This method can be attributed to Pearson \cite{pearsonMM} and it can be applied to any distribution for which there exists an unique relationship between its moments and parameters. In cases where the moments of a distribution can be mapped to the distributional parameters, the MM uses a Gaussian approximation to the moments to provide a closed form parameters estimation. The MM estimation is known to exist for the Gamma distribution, the Beta distribution or the Pareto distribution \footnote{http://www.math.uah.edu/stat/point/Moments.html}. 
Another classic approach for parameter estimation is the well known maximum likelihood (ML), based in the maximization of the data log-likelihood; historically it is known to have been used by Gauss or Laplace amongst others \cite{oldest_ML} but it was after the 1950s with the appearance of modern computers that it became relevant in practice; Fisher is believed to have been the first one using ML in the computational field. While the MM and ML approaches provide point estimates for the distribution parameter values, a full Bayesian estimation approach introduces distributions over the parameters \cite{MB}; when choosing conjugate prior distributions over the parameters \cite{MB,bishop}, the posterior expectations can be analytically derived.

Surprisingly, as far as we know, the method of moments estimation for the Inverse Gamma is not present in the literature. Further, it is well known that one can find a ML solution for the Inverse Gamma scale parameter for a given shape parameter. However, it is not possible to find a ML closed form solution for the shape parameter and in practice numerical optimization is required \cite{Lit:invg1,Lit:invg2}. Regarding the Bayesian estimation, the conjugate prior for the scale parameter is known to be Gamma distributed \cite{Fink97acompendium} but there is no conjugate prior for the shape parameter present in the literature. 
  
In section \ref{mt1} we introduce all the methods considered in this article. In section \ref{MM} we introduce the Method of Moments, in section \ref{ML} two novel Maximum Likelihood algorithms and in section \ref{bayes} we introduce two Bayesian parameter estimation strategies; section \ref{b:sc} reviews the methodology to find the scale posterior expectation, in section \ref{b:sh} we introduce a novel conjugate prior for the shape parameter that allows for the first time the analytic shape posterior estimation; in section \ref{IG_shape_post2} we introduce a Bayesian algorithm based on likelihood approximation and a corresponding conjugate prior. To be able to asses the quality of each algorithm, in section \ref{sec:KLIG} we introduce the KL-divergence between two Inverse Gamma distributions. 
Then, in section \ref{results} we provide some results. Section \ref{res:st} provides a theoretical comparison between the presented ML and Bayesian strategies and section \ref{res:sh} provides a visualization of different hyper parameters settings for the first shape conjugate prior.  In section \ref{num_res} we provide a numerical comparison between the five algorithms; in section \ref{num_res1} we quantify the quality of the algorithms in terms of KL divergences with respect to the true distributions and in section \ref{num_res2} we address the bias introduced by the estimators. We conclude the paper with a brief discussion in section \ref{dis}.

\section{Methods}
\label{mt1}
Consider the Inverse Gamma distribution
\begin{align}
\label{ig}
\mathcal{IG}(x|\alpha,\beta)= \frac{\beta^{\alpha} x^{- \alpha -1}}{\Gamma(\alpha)} \exp(\frac{-\beta}{x}),
\end{align} 
defined 
for any positive real parameter values $\alpha$ and $\beta$, denoting the distribution shape and scale respectively, and where $\Gamma$ is the Gamma function. The Inverse Gamma  is defined over the support $x \in \mathbb{R}^+$ and its first two moments are \cite{cook2008inverse}
\begin{align}
\label{moments}
 \mathbb{E}_{\mathcal{IG}}[x]=\frac{\beta}{\alpha -1}    \hspace{2 cm} \mathbb{E}_{\mathcal{IG}}[(x-\mathbb{E}_{\mathcal{IG}}[x])^2]=\frac{\beta^2}{(\alpha-1)^2(\alpha-2)}. 
\end{align}

Given a vector of positive real values $\textbf{x}=\{x_1,\ldots,x_n \}$, in the following subsections we present different algorithms to find posterior estimations $\hat{\alpha}$ and $\hat{\beta}$ for the distribution parameters $\alpha$ and $\beta$. 
\subsection{Method of Moments (MM)}
\label{MM}

Using a Gaussian approximation to the moments of the Inverse Gamma distribution (\ref{moments}) we obtain the following:
\begin{align*}
\mu \approx \frac{\beta}{\alpha-1},  \hspace{2cm} v \approx \frac{\beta^2}{(\alpha-1)^2 (\alpha-2)}, 
\end{align*}
where $\mu$ and $v$ are the mean and variance estimated from the observed data vector $\textbf{x}=\{x_1,\ldots,x_n \}$.
Solving this system for $\alpha$ and $\beta$ we obtain
\begin{align}
\label{mmig}
\hat{\alpha} \approx  \frac{\mu^2}{v}+2,  \hspace{2 cm} \hat{\beta} \approx \mu (\frac{\mu^2}{v}+1). 
\end{align}
These closed form solutions form the basis of the MM estimation for the parameters of the Inverse Gamma distribution.
Algorithm 1 summarizes the process.
\begin{algorithm}
\label{a4}
\caption{MM for Inverse Gamma}
\begin{algorithmic}
\REQUIRE $\textbf{x}=\{x_1,\ldots,x_n \}, x_i > 0$ 
\STATE  $\mu=\frac{1}{n} \sum_{i=1}^{n} x_i$ 
\STATE  $v= \frac{1}{n-1} \sum_{i=1}^{n} (x_i - \mu)^2$
\STATE $\hat{\alpha} = \frac{\mu^2}{v}+2$ 
\STATE $\hat{\beta} = \mu ((\frac{\mu^2}{v})+1)$ 
\RETURN $\hat{\alpha}$,$\hat{\beta}$
\end{algorithmic}
\end{algorithm}

\subsection{Maximum Likelihood (ML)}
\label{ML}
The log-likelihood of the positive vector of observations $\textbf{x}= \{x_1,\ldots,x_n \}$ under the Inverse Gamma distribution (\ref{ig}) can be written as
\begin{align}
\label{igll}
\log \mathcal{IG}(\textbf{x}|\alpha,\beta) = - n (\alpha+1) \overline{\log\textbf{x}} - n \log \Gamma(\alpha)  + n \alpha \log \beta   - \sum_{i=1}^{n}\beta x_i^{-1}
\end{align}
\normalsize
where the upper bar operand indicates the arithmetic mean. Finding ML estimations for the parameters is achieved by maximizing 
equation (\ref{igll}) with respect to the parameters $\{\alpha,\beta \}$; it is easy to verify that equation (\ref{igll}) has a maximum at 
\begin{align}
\label{thML}
{\beta} = \frac{ n\alpha} { \sum_{i=1}^{n} x_i^{-1}},
\end{align}
and that direct maximization of equation (\ref{igll}) with respect to $\alpha$ it is not possible. Substituting equation (\ref{thML})  into equation (\ref{igll}) and noting that $\sum_{i=1}^{n} \frac{1}{ x_i \sum_{j=1}^{n} x_j^{-1}}=1$ gives
\begin{align*}
\log \mathcal{IG}(\textbf{x}|\alpha) =  - n (\alpha+1) \overline{\log\textbf{x}} - n \log \Gamma(\alpha)  + 
n \alpha \log \alpha +
\end{align*}
\begin{align}
\label{igll3}
+n \alpha \log n- n \alpha \log \sum_{i=1}^{n} x_i^{-1} - n\alpha
\end{align}
\normalsize
Direct maximization of equation (\ref{igll3}) with respect to $\alpha$ is (obviously) also not possible. Using the linear constrain 
\begin{align}
\label{subst}
\alpha \log (\alpha) \ge (1+\log \alpha_0)(\alpha-\alpha_0) + \alpha_0 \log(\alpha_0)
\end{align}
and substituting (\ref{subst}) into equation (\ref{igll3}) provides a lower bound on the log-likelihood. Differentiating with respect to $\alpha$, equaling to zero and solving for $\alpha$ results in 
\begin{align}
\label{alphait}
\alpha= \Psi^{-1} \left( \log n \alpha_0 -\log\sum_{i=1}^{n} x_i^{-1} - \overline{\log{\textbf{x}}}  \right).
\end{align}
where $\Psi$ represents the \emph{digamma function}.

Using the MM relationship for the shape parameter (equation \ref{mmig}, left side) to initialize $\alpha_0$ in (\ref{alphait}), and iteratively updating $\alpha_0$ by $\alpha$ till convergence, we get $\hat{\alpha}$; we can then use equation ($\ref{thML}$) to get $\hat{\beta}$. Algorithm 2 summarizes the process:
\begin{algorithm}
\label{a5}
\caption{ML1}
\begin{algorithmic}
\REQUIRE $\textbf{x}=\{x_1,\ldots,x_n \}, x_i > 0$ 
\STATE  $\mu= \bar{x} = \frac{1}{n} \sum_{i=1}^{n} x_i$
\STATE  $v= \frac{1}{n-1} \sum_{i=1}^{n} (x_i - \mu)^2$
\STATE $\alpha$ = $\frac{\mu^2}{v}+2$ 
\STATE $C= -\log\sum_{i=1}^{n} x_i^{-1} - \frac{1}{n} \sum_{i=1}^{n}\log{x_i}$
\REPEAT
\small
\STATE ${\alpha}$ $\leftarrow$ $\Psi^{-1}(  \log n \alpha +C )$
\normalsize
\UNTIL {convergence}
\STATE $\hat{\alpha} = \alpha$
\STATE $\hat{\beta}$ = $\frac{n \hat{\alpha}}{\sum_{i=1}^{n} x_i^{-1} }$ 
\RETURN $\hat{\alpha}$,$\hat{\beta}$
\end{algorithmic}
\end{algorithm}

Inspired by \cite{Minka_newton,minka2002estimating} we develop next another algorithm for ML estimation of the Inverse Gamma parameters. We approximate to the Inverse Gamma log-likelihood presented in equation (\ref{igll3}) by
\begin{align}
\label{approx}
f(\alpha) = k_0 + k_1 \alpha + k_2 \log \alpha.
\end{align}
Taking first and second derivatives of $f(\alpha)$
and matching $f(\alpha)$ and its derivatives to $\log \mathcal{IG}(\textbf{x}|\alpha)$ and its first two derivatives with respect to $\alpha$ respectively, we obtain values for $k_0,k_1,k_2$, namely  
\begin{align}
\label{k2}
k_2 = n \big( \alpha^2 \Psi'(\alpha) - \alpha \big)
\end{align}
\begin{align}
\label{k1}
k_1= n \big( \overline{-\log \textbf{x}} - \Psi(\alpha) + \log n\alpha -\log \sum_{i=1}^{n}x_i^{-1} - \alpha \Psi'(\alpha) +1\big)
\end{align}
\begin{align}
\label{k1}
k_0= \log IG (x|\alpha) - k_1 \alpha - k_2 \log \alpha 
\end{align}
with $\log IG (\textbf{x}|\alpha)$ as given in equation (\ref{igll3}).

If $f''(\alpha) < 0$, $f(\alpha)$ has a maximum at $\hat{\alpha}$ iff $f'(\hat{\alpha})=0$ iff $\hat{\alpha}=\frac{-k_2}{k_1}$. This process gives the following update rule
\begin{align}
\frac{1}{\hat{\alpha}}=\frac{1}{\alpha} + \frac{-\overline{\log \textbf{x}} - \psi(\alpha) + \log n\alpha - \log \sum_{i=1}^{n} x_i^{-1}}{\alpha^2 (\frac{1}{\alpha}-\psi'(\alpha))} . 
\end{align}

As before we use the MM relationship for the shape parameter (equation \ref{mmig}, left side) to initialize $\alpha$. After iteration till convergence to get $\hat{\alpha}$ we use equation ($\ref{thML}$) to get $\hat{\beta}$. Algorithm 3 summarizes the process:
\begin{algorithm}
\label{a5}
\caption{ML2}
\begin{algorithmic}
\REQUIRE $\textbf{x}=\{x_1,\ldots,x_n \}, x_i > 0$ 
\STATE  $\mu= \bar{x} = \frac{1}{n} \sum_{i=1}^{n} x_i$
\STATE  $v= \frac{1}{n-1} \sum_{i=1}^{n} (x_i - \mu)^2$
\STATE $\alpha$ = $\frac{\mu^2}{v}+2$ 
\STATE $C= -\log\sum_{i=1}^{n} x_i^{-1} - \frac{1}{n} \sum_{i=1}^{n}\log{x_i}$
\REPEAT
\small
\STATE $\frac{1}{{\alpha}}=\frac{1}{\alpha} + \frac{C - \psi(\alpha) + \log n\alpha}{\alpha^2 (\frac{1}{\alpha}-\psi'(\alpha))} $ 
\normalsize
\UNTIL {convergence}
\STATE $\hat{\alpha} = \alpha$
\STATE $\hat{\beta}$ = $\frac{n \hat{\alpha}}{\sum_{i=1}^{n} x_i^{-1} }$ 
\RETURN $\hat{\alpha}$,$\hat{\beta}$
\end{algorithmic}
\end{algorithm}

\subsection{Bayesian Learning (BL)}
\label{bayes}
Using Bayes rule to find the posterior probability of the Inverse Gamma parameters we have that
\begin{align}
\label{by}
p(\theta|\textbf{x}) = \frac{ p(\textbf{x}|\theta) p(\theta)}{p(\textbf{x})},
\end{align} 
where $\theta=\{\alpha, \beta \}$ and $\textbf{x}= \{x_1,\ldots,x_n \}$ is some positive vector of observations. Since the denominator only depends on data the posterior is proportional to the likelihood multiplied by the prior
\begin{align}
\label{by}
p(\theta|\textbf{x}) \propto  p(\textbf{x}|\theta) p(\theta).
\end{align} 
Obtaining analytical solutions for the parameters $\theta$ requires the use of conjugate priors \cite{bishop,Fink97acompendium}. A prior is called conjugate with a likelihood function if the prior functional form remains unchanged after multiplication by the likelihood. 
Remember here that the likelihood of $\textbf{x}= \{x_1,\ldots,x_n \}$ under the Inverse Gamma distribution is
\begin{align}
\label{iglk}
p(\textbf{x}|\theta )=\mathcal{IG}(\textbf{x}|\alpha,\beta)= \prod_{i=1}^{n} \frac{\beta^{\alpha} x_i^{- \alpha -1}}{\Gamma(\alpha)} \exp\left(\frac{-\beta}{x_i}\right).
\end{align} 

Next, in subsection \ref{b:sc}, we review the conjugate prior for the scale parameter and in subsection \ref{b:sh} we introduce the one missing in the literature, the conjugate prior for the shape parameter.  

\subsubsection{Scale parameter $\beta$} 
\label{b:sc}
A well known conjugate prior for $\beta$, the scale parameter of the Inverse Gamma distribution, is a Gamma distribution parametrized using shape $d$ and rate $e$ \cite{Fink97acompendium},  
\begin{align}
\label{pb}
p (\beta)=  \mathcal{G}(\beta|d,e)= \frac{e^d \beta^{d-1}}{\Gamma(d)} \exp (-e \beta).
\end{align}

For completeness, in the remaining of this subsection we show that equation (\ref{pb}) is conjugate to the Inverse Gamma likelihood and provide the posterior expectation on $\beta$. Given the observations vector $\textbf{x}$, we multiply the Inverse Gamma data likelihood (\ref{iglk}) by the prior on scale (\ref{pb}) to get its posterior, $q(\beta)$
\begin{align}
q(\beta) \propto \left( \prod_{i=1}^{n} \frac{\beta^{\alpha} x_i^{- \alpha -1}}{\Gamma(\alpha)} \exp\left(\frac{-\beta}{x_i}\right) \right)
\frac{e^d \beta^{d-1}}{\Gamma(d)} \exp (-e \beta)
\end{align} 
Keeping only $\beta$ dependent terms
\begin{align*}
q(\beta) \propto 
e^d \beta^{d-1+n\alpha} \exp \left( -\beta (e  + \sum_{i=1}^{n} x_i^{-1}) \right), 
\end{align*} 
so the posterior is Gamma distributed $q(\beta) = \mathcal{G}(\beta|\hat{d},\hat{e})$ with parameter values
\begin{align}
\label{td}
\hat{d}= d + n \alpha,    \hspace{2cm} \hat{e}=e+\sum_{i=1}^{n} x_i^{-1}.
\end{align} 
Since $\beta$ is Gamma distributed its posterior expectation is
\begin{align}
\label{beta}
\hat{\beta} =  \frac{\hat{d}}{\hat{e}}.
\end{align} 

\subsubsection{Shape parameter $\alpha$ (BL1)}
\label{b:sh}
Inspired by the unnormalized prior used for the shape parameter of the gamma distribution \cite{Fink97acompendium}, we propose here an unnormalized prior
for the shape parameter $\alpha$ of the inverse gamma distribution of the form
\begin{align}
\label{IG_shape_prior}
p (\alpha) \propto \frac  {a^{-\alpha-1} \beta^{\alpha c}}   {\Gamma(\alpha)^{b}},
\end{align}
where $\beta$ is the Inverse Gamma scale parameter and $\{a,b,c\} \in \mathcal{R}^{+}$ are hyper parameters. 
Given some observations $\textbf{x}$, we multiply the associated likelihood under the Inverse Gamma distribution (\ref{iglk}) by the proposed prior on shape (\ref{IG_shape_prior}) to obtain an expression for the posterior distribution $q(\alpha)$
\begin{align*}
q(\alpha) \propto \left( \prod_{i=1}^{n} \frac{\beta^{\alpha} x_i^{- \alpha -1}}{\Gamma(\alpha)} \exp(\frac{-\beta}{x_i}) \right)
\frac  {a^{-\alpha-1} \beta^{\alpha c}}   {\Gamma(\alpha)^{b}}.
\end{align*} 
Keeping only $\alpha$ dependent terms we note that 
\begin{align}
\label{IG_shape_post}
q (\alpha) \propto \frac  {\hat{a}^{-\alpha-1} \beta^{\alpha \hat{c}}}   {\Gamma(\alpha)^{\hat{b}}}
\end{align}
with 
\begin{align}
\label{ha}
\hat{a}= a \prod_{i=1}^{n} x_i, \hspace{2cm} \hat{b}=b+n, \hspace{2cm} \hat{c}= c + n. 
\end{align} 
This proves that equation (\ref{IG_shape_prior}) is a conjugate prior for the Inverse Gamma shape. Finding the posterior expectation of $\alpha$, $\hat{\alpha}$, implies computing the expectation of $q(\alpha)$. Here we use the Laplace approximation to (\ref{IG_shape_prior}), which can be shown to be a Gaussian with mean 
\begin{align*}
m= \Psi^{-1}\Big(\frac{-\log a + c \log \beta}{b}\Big),
\end{align*}
and precision $b\Psi_1(m)$, where $\Psi_{1}(\alpha) =\frac{d \Psi(\alpha)}{d\alpha}$.
Consequently we approximate the expected shape by
\begin{align}
\label{mu1}
\hat{\alpha} \approx \Psi^{-1}\Big(\frac{-\log \hat{a} + \hat{c} \log \beta}{\hat{b}}\Big).
\end{align}
We note here that the expectation of the Laplace approximation to $q(\alpha)$ corresponds to the maximum a posteriori (MAP) estimate of $q(\alpha)$. The use of the Laplace approximation in this context then reduces to using a MAP estimation in place of the expected value. 
Note also that in order to compute the expectation (\ref{mu1}) we need to estimate $\log \hat{a}$ 
\begin{align}
\label{lha}
\log \hat{a}= \log a + \sum_{i=1}^{n} \log x_i,
\end{align} 
and not longer require $\hat{a}$ in equation (\ref{ha}). This fact has the advantage of avoiding numerical issues for large sample sizes. Further, substituting $\hat{\beta}= \frac{\hat{d}}{\hat{e}}=\frac{d + n\alpha}{e+\sum_{i=1}^{n} x_i^{-1}}$ into equation (\ref{mu}) we obtain an expression with no $\beta$ dependence which is more compact for algorithmic use, namely
\begin{align}
\label{mu2}
\hat{\alpha} \approx \Psi^{-1}\Big(\frac{-\log \hat{a} + \hat{c} \big( \log (d + n\alpha) - \log (e+\sum_{i=1}^{n} x_i^{-1}) \big)}{\hat{b}}\Big).
\end{align}

\begin{algorithm}
\label{aa100}
\caption{BL1}
\begin{algorithmic}
\REQUIRE $\textbf{x}=\{x_1,\ldots,x_n \}, x_i > 0$ and $\{a, b, c, d, e\}$ 
\STATE  $\mu= \frac{1}{n} \sum_{i=1}^{n} x_i$
\STATE  $v= \frac{1}{n-1} \sum_{i=1}^{n} (x_i - \mu)^2$
\STATE $\alpha = \frac{\mu^2}{v}+2$
\STATE $\hat{e}=e+\sum_{i=1}^{n} x_i^{-1}$
\STATE  $\log \hat{a}= \log a + \sum_{i=1}^{n} \log x_i$
\STATE $\hat{b}=b+n$
\STATE $\hat{c}= c + n$
\STATE $C=\log (\hat{e})$
\REPEAT
\STATE $\alpha \leftarrow \Psi^{-1}\Big(\frac{-\log \hat{a} + \hat{c} \big( \log (d + n \alpha) - C \big)}{\hat{b}}\Big)$
\UNTIL {convergence}
\STATE $\hat{\alpha}=\alpha$
\STATE $\hat{d}= d + n \hat{\alpha} $
\STATE $\hat{\beta}=\frac{\hat{d}}{\hat{e}}$
\RETURN $\hat{\alpha}$,$\hat{\beta}$
\end{algorithmic}
\end{algorithm}
Algorithm 4 summarizes the Bayesian approach for learning the Inverse Gamma parameters. First, the MM algorithm is used to initialize $\alpha$ (equation (\ref{mmig}) left panel); $\hat{e}$, $\log \hat{a}$, $\hat{b}$ and $\hat{c}$ are computed through equations (\ref{td} right side), (\ref{lha}) and (\ref{ha}). Then equation (\ref{mu2}) is iterated till convergence to obtain the expected $\hat{\alpha}$. Finally $\hat{d}$ is computed using equation (\ref{td} left side) and 
$\hat{\beta}$ through  (\ref{beta}).

\subsubsection{Shape parameter $\alpha$ with function approximation (BL2)}
\label{IG_shape_post2}
Given an initial parameter value estimation $\alpha$, one can use the approximation to the log-likelihood presented in equation ($\ref{approx}$). As previously explained, learning the parameters $k_0,k_1,k_2$ is straightforward. We can now define a prior on $\alpha$ that is conjugate with this approximated log-likelihood, namely
\begin{align}
\label{log_cp}
\log p(\alpha) \propto w_0 + w_1 \alpha + w_2 \log \alpha
\end{align} 
for any set of hyper parameter values $w_0,w_1,w_2$. It is straightforward to observe that the posterior values for the hyper parameters are 
\begin{align}
\tilde{w}_i=w_i +k_i, \hspace{1cm}  \forall i \in \{0,1,2 \}.
\end{align} 
Once more we use the Laplace approximation to (\ref{log_cp}), which can be shown to be a Gaussian with mean 
\begin{align*}
m= \frac{-\tilde{w_2}}{\tilde{w_1}}
\end{align*}
and precision $\frac{\tilde{w_2}}{\alpha^2}$, 
Consequently we approximate the expected shape by
\begin{align}
\label{mu}
\hat{\alpha} \approx \frac{-\tilde{w_2}}{\tilde{w_1}}.
\end{align}
Again, the Laplace approximation corresponds to the maximum a posteriori (MAP) estimate. Algorithm 5 summarizes the process.

\begin{algorithm}
\label{aa1000}
\caption{BL2}
\begin{algorithmic}
\REQUIRE $\textbf{x}=\{x_1,\ldots,x_n \}, x_i > 0$ and $\{w_1,w_2\}$ 
\STATE  $\mu= \frac{1}{n} \sum_{i=1}^{n} x_i$
\STATE  $v= \frac{1}{n-1} \sum_{i=1}^{n} (x_i - \mu)^2$
\STATE $\alpha = \frac{\mu^2}{v} +2$

\REPEAT
\STATE $k_1= n \big( \overline{-\log \textbf{x}} - \Psi(\alpha) + \log n\alpha -\log \sum_{i=1}^{n}x_i^{-1} - \alpha \Psi'(\alpha) +1\big)$
\STATE $k_2= n \big( \alpha^2 \Psi'(\alpha) - \alpha \big)$ 
\STATE $\tilde{w_1}= w_1+k_1$
\STATE  $\tilde{w_2}= w_2+k_2$
\STATE $\alpha \leftarrow \frac{-\tilde{w_2}}{\tilde{w1}}$
\UNTIL {convergence}
\STATE $\hat{\alpha}=\alpha$
\STATE $\hat{d}= d + n \hat{\alpha} $
\STATE $\hat{\beta}=\frac{\hat{d}}{\hat{e}}$
\RETURN $\hat{\alpha}$,$\hat{\beta}$
\end{algorithmic}
\end{algorithm}

\subsection{KL-divergence}
\label{sec:KLIG}
Let $p(x)=IG(x|\alpha,\beta)$ and $q(x)=IG(x|\hat{\alpha},\hat{\beta})$ be two Inverse Gamma distributions with parameters $\{\alpha,\beta\}$ and $\{\hat{\alpha},\hat{\beta}\}$ respectively. Then the KL-divergence between $p$ and $q$ is
\begin{align}
KL[p||q]= (\alpha - \hat{\alpha}) \Psi(\alpha) + \hat{\beta} (\frac{\alpha}{\beta}) - \alpha + \log \frac{\beta^{\hat{\alpha}+1} \Gamma(\hat{\alpha})}{ \beta \hat{\beta}^{\hat{\alpha}} \Gamma(\alpha)}. 
\label{KLIG}
\end{align}
The proof is straight forward using the definition of KL divergence between the two distributions
\begin{align}
KL[p||q] = \int_{0}^{\infty} p(x) \log \frac{p(x)}{q(x)} dx
\end{align}
and requires the use of 
\begin{align}
\label{pdf}
\int_{0}^{\infty} p(x) dx =1,
\end{align}
\begin{align}
\label{logp}
\mathbb{E}_{p}[\log p(x)] = (1+\alpha) \Psi(\alpha) - \alpha - \log (\beta \Gamma(\alpha)),
\end{align}
\begin{align}
\label{logx}
\mathbb{E}_{p}[\log x] = \log(\beta) - \Psi(\alpha),
\end{align}
\begin{align}
\label{frac}
\mathbb{E}_{p}[x^{-1}] = \frac{\alpha}{\beta}.
\end{align}
Equation (\ref{pdf}) is a direct consequence of $p(x)$ being a pdf, and equations (\ref{logp},\ref{logx},\ref{frac}) can be easily derived from the definition of expectation of a function under a pdf
\begin{align}
\mathbb{E}_{p}[f(x)] = \int_{0}^{\infty} p(x) f(x) dx.
\end{align}

\section{Results}
\label{results}
In section \ref{res:st} we derive the relationships between the ML with the BL algorithms presented. Then in section \ref{res:sh} we present examples showing the effect of different hyper parameters choices on the shape conjugate prior of the BL1 algorithm as well as demonstrating the methodology proposed to obtain the posterior expected shape in the Bayesian setting. In section \ref{num_res1} we present some numerical results comparing the five different algorithms presented in terms of KL divergences and in section \ref{num_res2} we numerically study the bias introduced by the different estimators.

\subsection{Relation between ML and BL algorithms}
\label{res:st}
Note first that both maximum likelihood algorithms have the same fix point at $\alpha$ such that $\psi(\alpha)= \log n\alpha -\log \sum_{i=1}^{n} x_i^{-1} -\overline{\log{\textbf{x}}}$; consequently they will both provide the same solution. Furthermore, both ML and BL algorithms have notable similarities; first, note that the ML and the Bayesian scale estimators presented are 
\begin{align}
\hat{\beta}_{ML}= \frac{n \alpha}{\sum_{i=1}^{n} x_i^{-1} } \hspace{2cm} \hat{\beta}_{BL}= \frac{\hat{d}}{\hat{e}}=\frac{d + n\alpha}{e+\sum_{i=1}^{n} x_i^{-1}}
\end{align}
respectively. It is easy to observe that 
\begin{align}
\lim_{d=e \to 0} \hat{\beta}_{BL} \rightarrow \hat{\beta}_{ML},
\end{align}
which means that the Bayesian scale estimator tends to the ML one in the limit of an infinite variance gamma prior over $\beta$. 

With respect to the BL1 shape parameter estimation ($\alpha$), since initializing $b=c$ results in $\hat{b}=\hat{c}$, 
taking again the same limit we can rewrite the Bayesian $\alpha$ update as
\begin{align}
\lim_{ \substack{d=e \to 0 \\ b=c}} \hat{\alpha}_{BL} \rightarrow \Psi^{-1} \Big( \log n\alpha - \big( \log  \sum_{i=1}^{n} x_i^{-1}\big) -\frac{ \log a + \sum_{i=1}^{n} \log x_i}{\hat{b}}\Big).
\end{align}
Comparing this to the ML1 $\alpha$ update 
\begin{align}
\hat{\alpha}_{ML1} = \Psi^{-1}(  \log n \alpha - \big( \log \sum_{i=1}^{n} x_i^{-1} \big) - \overline{\log{\textbf{x}}} ),
\end{align}
it is clear that both models perform an iterative update of the shape parameter dependent in the inverse of the gamma function of $\log n \alpha$ plus a data dependent constant which differs for ML1 and BL1; they also have the same dependence on one of the data dependent terms, $\log \sum_{i=1}^{n} x_{i}^{-1}$, independently of the hyper parameter values used for the prior on $\alpha$ (under the assumption b=c). Further, since $\hat{b}=b+n$, in the extreme case of considering a very small hyper parameter value for $b$, we can simplify the Bayesian  estimation update further, resulting in 
\begin{align}
\lim_{ \substack{d=e \to 0 \\ b=c \to 0}} \hat{\alpha}_{BL} \rightarrow \Psi^{-1} \Big( \log n\alpha - \big( \log  \sum_{i=1}^{n} x_i^{-1}\big) - \overline{\log \textbf{x}} -\frac{\log a}{n}  \Big).
\end{align}
Remembering that the posterior estimation for $b$ and $c$ are $\hat{b}=b+n$ and $\hat{c}=c+n$, this is a very interesting result that shows that when choosing little informative hyper parameter values for $b$ and $c$, the Bayesian estimation involves a small sample bias correction term $\frac{-\log a}{n}$ which tends to $0$ as the number of observed samples increases. Clearly, relating the shape update of BL1 and that of ML2 is not so straightforward. The possibility undercovering such relationship might have been lost due to the use of the Laplace approximation to the posterior of $\alpha$ and its investigation is focus of current research. There is nevertheless by construction a close relationship between ML2 and BL2; in fact they are equivalent when considering a flat prior over $\alpha$, that is when $w_1=w_2=0$ in equation (\ref{log_cp}).

\subsection{Shape conjugate priors hyper parameters}
\label{res:sh}
\begin{figure}[!]
\centering  	
\includegraphics[width=.8\textwidth]{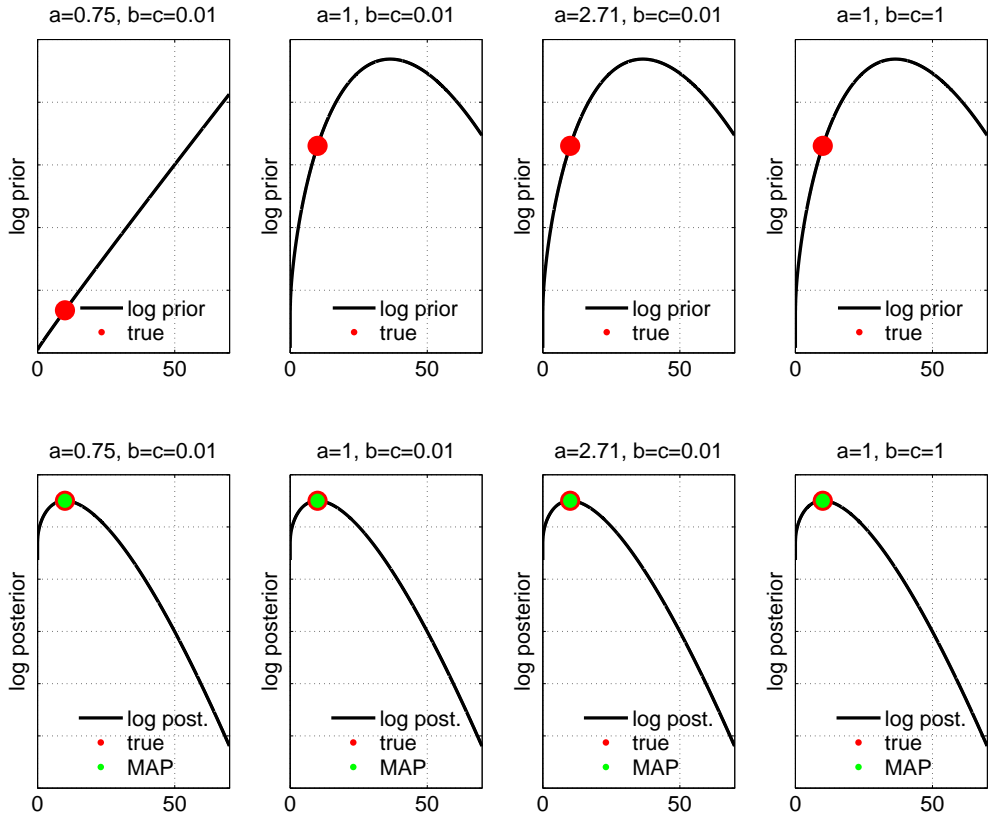}.
\caption{Top row shows the log-prior $\log p(\alpha|a,b,c,\beta)$ and the bottom row the log-posterior $\log p(\alpha|\hat{a},\hat{b},\hat{c},\hat{\beta})$. The red dots mark the shape parameter $\alpha$ from which data was generated. The green dots represent the Bayesian posterior estimation.} 
\label{fig1}
\end{figure}

When choosing hyper parameters for a prior, non-informative (flat) priors are usually preferred to get unbiased estimators. A flat prior for the shape parameter in the BL2 case can easily be achieved by choosing $w_1=w_2=0$ in equation (\ref{log_cp}). For simplicity in the remaining of this work we consider fixed values for the hyper parameters of the BL2 algorithm as $w_0=1$, $w_1=w_2=0$. On the other side, looking at equation (\ref{IG_shape_prior}) it is clear that choosing a flat prior for BL1 is not straightforward. To get some intuition we provide next some numerical examples illustrating such prior for different hyper parameter values and demonstrating the behavior of the posterior hyper parameter updates (equations  \ref{ha},\ref{lha}) as well as that of the Laplace approximation (equation \ref{mu}) used to obtain the expected shape $\hat{\alpha}$ under different hyper parameter values choices in the BL1 algorithm. We generate 1000 samples from an Inverse Gamma distribution with parameter values $\alpha=10$ and $\beta=25$. $\hat{\beta}$ is computed through equation (\ref{mmig}) left panel followed by equations (\ref{td},\ref{beta}) where a flat prior was used, namely $d=e=0.01$. For the shape prior hyper parameters we consider different values $a,b,c$, and $\hat{\alpha}$ is computed independently on each case through equations (\ref{ha},\ref{lha},\ref{mu}). In Figure \ref{fig1}, the top row shows the log-prior $\log p(\alpha|a,b,c,\hat{\beta})$ for the parameter values $a,b,c$ indicated in the titles as a function of the shape value $\alpha$ represented in the x-axis. The bottom row shows the corresponding log-posteriors $\log p(\alpha|\hat{a},\hat{b},\hat{c},\hat{\beta})$. In all cases the red dot represents the true shape value, $\alpha=10$, and the green ones in the bottom row represents the posterior estimated value $\hat{\alpha}$.
The log-posterior is much sharper around the correct value. The expectation we obtained for the posterior of $\alpha$, $\hat{\alpha}$ marked as as green circle, is a very good approximation independently of the chosen hyper parameter values. Since the parameter value for the scale $\beta$ was also estimated using Bayesian inference and its part of the prior and posterior on $\alpha$, the figure indirectly confirms also the proper behaviour of the scale Bayesian estimation procedure. These examples are representative of the general behaviour observed for a broad range of hyper parameter values $\{a,b,c\}$.


\subsection{Numerical Results}
\label{num_res}

\subsubsection{Numerical algorithms comparison}
\label{num_res1}

In the remainder of this section we perform a numerical comparison between the method of moments (MM), the maximum likelihood (ML1 and ML2) and the Bayesian approaches (BL1 and BL2). For each simulation we generate varying amount of $N$ samples from a Inverse Gamma distribution with fixed parameters $\alpha$ and $\beta$ and we applied the five presented algorithms for estimating the parameters. The parameter values $\alpha$ and $\beta$ are initialized to different random positive numbers at each simulation. For each $N$ we performed 500 different simulations. The ML and the BL algorithms are considered to converge when the relative $\alpha$ parameter change between two consecutive iterations is smaller than $10^-6$. In this example the hyper parameter values are fixed to $a=1$, $b=c=0.01$ and $d=e=0.01$.
\begin{figure}[!]
\centering  	
\includegraphics[width=.9\textwidth]{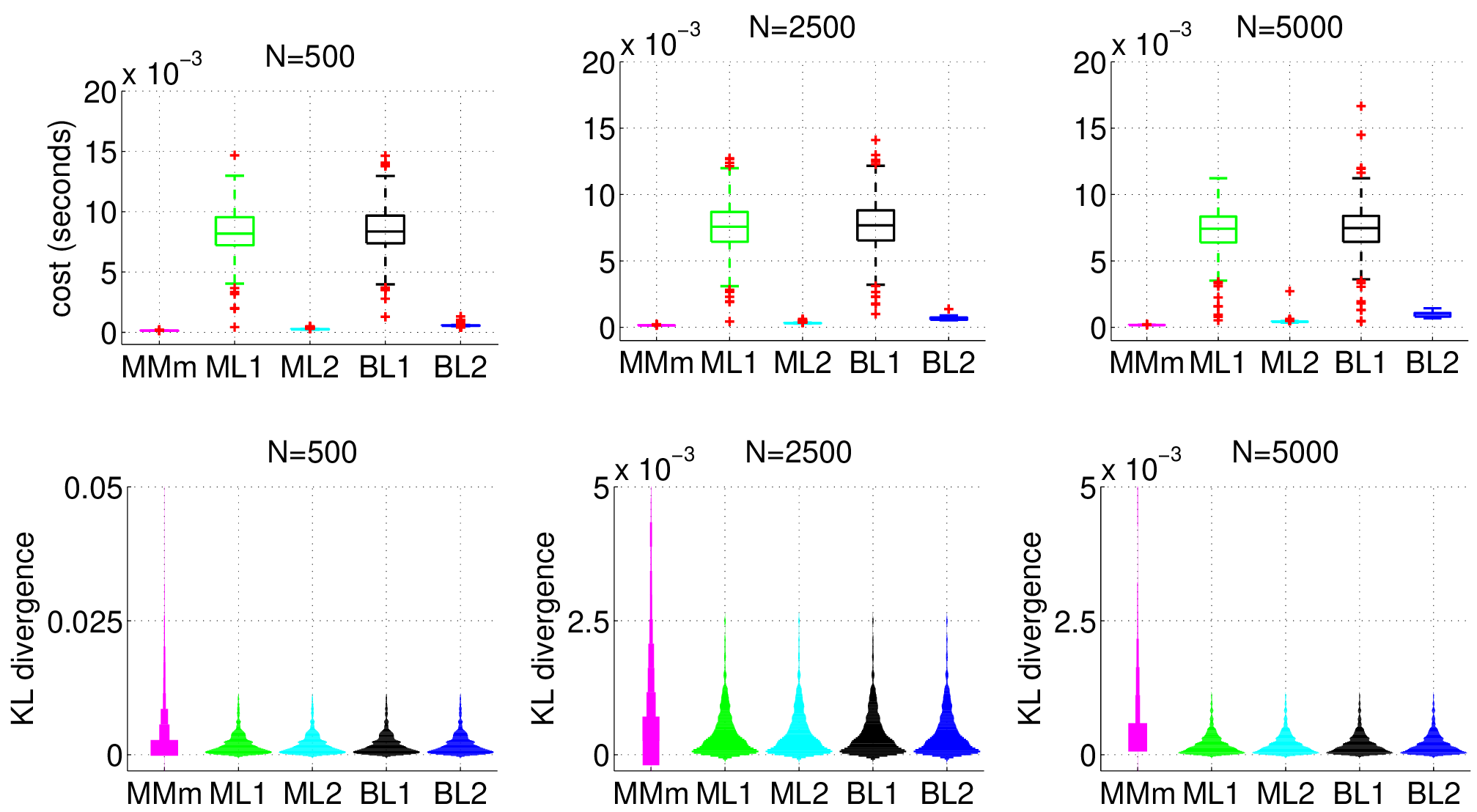}.
\caption{The top row shows violin plots of the KL divergences between the true and the estimated distributions for each of the 5 algorithms as indicated on the x-axis and for different  number of observed samples ($N$) on each column. The different algorithms are also color coded for better visualization. The bottom row shows box plots of the computational time required by the different algorithms expressed in seconds.} 
\label{fig:err}
\end{figure} 
The top row of Figure \ref{fig:err} shows violin plots of the KL-divergences between the true and the estimated distributions for each of the 5 algorithms and for different number of observed samples ($N$) as indicated on each column. The method of moments (MM) shows larger deviations from the true distribution, while the ML and the BL algorithms provide very similar solutions. To asses statistical differences between the error obtained by the different algorithms we used a Wilcoxon rank test between each pair of models at each $N$ independently and assess statistical significance at values $p < 0.01$. Independently of the number of samples, the MM algorithm is always significantly worse than the other two. The ML and the BL algorithms were not found to be significantly different. The bottom row shows box plots of the computational costs in seconds; the mean, $25^{-th}$ and $75^{-th}$ percentile are presented on each box and the red crosses represent outliers of the estimated distributions. Obviously the ML1 and the BL1 approaches, as presented here, are more expensive than MM since they require iteration and in fact initialize with MM. Remark here the computational advantage of ML2 and BL2 with respect to ML1 and the BL1 algorithms; ML2 and BL2 required a mean of 4 iterations to converge.  The cost of ML1 and BL1 algorithms is equivalent which is not surprising given the theoretical comparison we performed previously. It is to note that, for both ML as well as the BL algorithms, the iterative process cost has no $N$ dependency inside the loop. Interestingly, we also observed that a single iteration of the ML or the BL algorithms already suffices to significantly improve with respect to the MM. 

Since on the basis of our analytic analyses we can conclude that small values for $b$ and $c$ are a reasonable choice, we kept fixed these at values $b=c=0.01$ and vary the values of $a$. We observed that the BL1 algorithm is no significantly different from the ML1 for a wide range of $a$ values. As a reference, values of $a$ in the range $[0.5, 5]$ were always optimal for simulations analogous to the one presented in figure \ref{fig:err}. Much higher or smaller values of $a$ resulted in better solutions than MM but worse than ML1; the difference with respect to ML1 was only observed at the smallest of the considered sample sizes, $N=500$ samples and vanished for the higher sample sizes. For the BL2 algorithm we restricted our analyses to the presented non informative prior.



\subsubsection{Numerical bias estimation}
\label{num_res2}
\begin{figure}[!]
\centering  	
\includegraphics[width=1\textwidth]{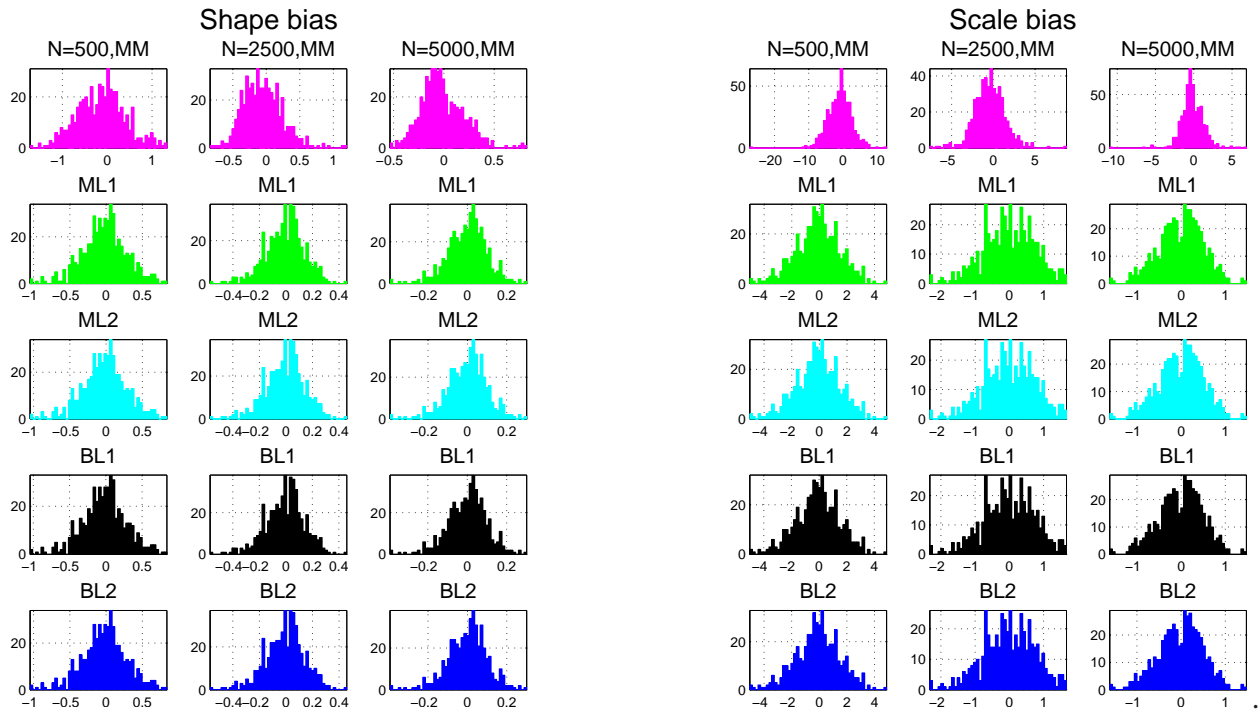}.
\caption{Histograms of the bias of the different estimators (rows, color coded), for each parameter (shape and scale) and different sample sizes $N=\{500,2500,5000 \}$} 
\label{fig:bias_hist}
\end{figure}

\begin{figure}[!]
\centering  	
\includegraphics[width=.9\textwidth]{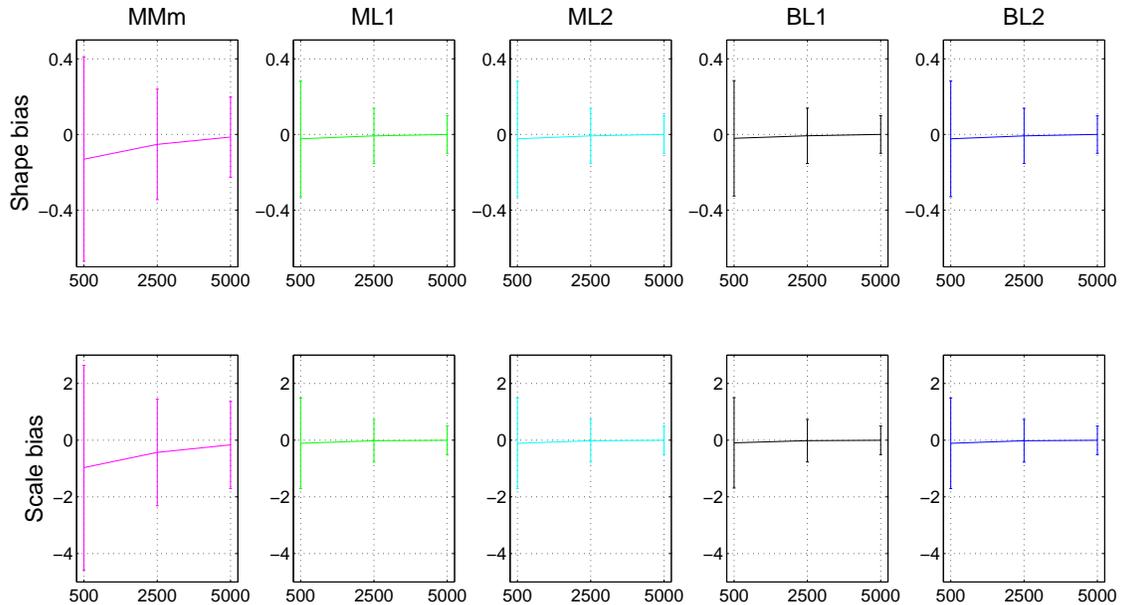}.
\caption{Mean and standard deviation of the bias in the estimation of shape and scale parameters (rows) for each of the considered estimators (columns) for the different samples sizes presented in the x-axis.} 
\label{fig:bias}
\end{figure}

In this section we numerically analyze the bias introduced by the estimators. 
In figure \ref{fig:bias_hist} we present histograms of the bias of the estimators for each parameter and each of the estimators (color coded) for the previously considered simulations. While both ML and BL algorithms bias distributions are approximetely Gaussians centered in zero, the MMm bias distributions are more skewed and provide also several outlayers. To quantify these differences, in figure \ref{fig:bias} we show error bars presenting the mean and standard derivation of such distributions. While the MMm provides more biased estimations, the ML and the BL algorithms show a small bias error with decresing variance for higher sample sizes.

\section{Discussion}
\label{dis}
We here presented 5 novel algorithms for the estimation of the parameters of a Inverse Gamma distribution and provide numerical evaluation of their estimation performance using synthetic data. The method of moments (MM) provides a closed form fast estimation; the maximum likelihood (ML) and the Bayesian approach (BL) use the method of moments as initialization and provide more accurate solutions, albeit at higher computational cost; ML2 and BL2 are much faster than ML1 and BL1 and converge in just a few iterations. Both ML algorithms converge to the same solution. BL2 converges also to the same solution when using a flat prior and we provided hyper parameter values for the new conjugate prior for the shape parameter used in BL1 for which the solution is also the same. Further study of the introduction of prior knowledge under the presented conjugate priors is focus of current research. We also showed numerically that the bias introduduced by the ML and the BL algorithms assymptotically tends to zero with increasing sample sizes. Here we reduced this analyses to a numerical approach due to intractability of a direct closed form bias estimation derivated from the appearance of the digamma function in the proposed shape estimators. Althought a more complex analyses (as in \cite{10.2307/1266892,RePEc:vic:vicewp:0908} for the Gamma distribution) could be possible and could result in corrections for the biases, this remains out of the scope of this work and is considered as ongoing research. 

The Bayesian Learning approaches introduced here has important implications since it can be used inside more complex Bayesian inference tasks, such as fitting mixture models involving Inverse Gamma distributions within a (Variational) Bayesian framework to be used for example in the context of medical image segmentation \cite{woolMM}. Extending this beyond modeling one dimensional data, the presented methodology can also be included in multivariate models, e.g. for Bayesian ICA or Non-negative matrix factorizations.  

\bibliographystyle{abbrv}
\bibliography{Gamma_InvGm.bib}

\end{document}